\documentclass[%
aps,
amsmath,amssymb,
superscriptaddress,
reprint%
]{revtex4-2}

\usepackage{graphicx}
\usepackage{dcolumn}
\usepackage{bm}

\usepackage{color}

\graphicspath{{./Figures_230703/}}

\usepackage{upgreek}
\usepackage{gensymb}
\usepackage{float}
\usepackage{xcolor}
\usepackage{physics}
\usepackage{longtable}
\usepackage{comment}

\usepackage{xr}
\makeatletter
\newcommand*{\addFileDependency}[1]{
	\typeout{(#1)}
	\@addtofilelist{#1}
	\IfFileExists{#1}{}{\typeout{No file #1.}}
}
\makeatother

\begin{document}
	
\title{On-chip all-electrical determination of the magnetoelastic coupling constant of magnetic heterostructures}

\author{Takuya Kawada}
\affiliation{Department of Physics, The University of Tokyo, Tokyo 113-0033, Japan}
\affiliation{Department of Physics, Graduate School of Science, Osaka University, Toyonaka, Osaka 560-0043, Japan}

\author{Isamu Yasuda}
\affiliation{Department of Physics, The University of Tokyo, Tokyo 113-0033, Japan}

\author{Masashi Kawaguchi}
\affiliation{Department of Physics, The University of Tokyo, Tokyo 113-0033, Japan}

\author{Masamitsu Hayashi}
\affiliation{Department of Physics, The University of Tokyo, Tokyo 113-0033, Japan}
\affiliation{Trans-scale quantum science institute, The University of Tokyo, Tokyo 113-0033, Japan}

\newif\iffigure
\figurefalse
\figuretrue

\date{\today}

\begin{abstract}
We have developed an approach to determine the magnetoelastic coupling constant of magnetic layers in thin film heterostructures.
The film is formed on a piezoelectric substrate between two interdigital transducers (IDT), a platform often used to construct a surface acoustic wave device.
With the substrate piezoelectricity, strain is induced into the film by applying a dc voltage to the IDTs.
The strain causes changes in the magnetization direction of the magnetic layer, which is probed by measuring changes, if any, in the transverse resistance of the heterostructure.
We find the extracted magnetoelastic coupling constant of the magnetic layer (CoFeB) depends on the film stacking. 
Such change can be accounted for provided that the elastic properties of the layers that constitute the heterostructures are taken into account.
The on-chip all-electrical approach described here provides a versatile means to quantitatively assess the magnetoelastic coupling constant of thin film heterostructures.
\end{abstract}

\maketitle
Magnonics\cite{Gulyaev2001,Puszkarski2003,Kruglyak2010,Barman2021} is a research field that focuses on the transport and manipulation of spin waves or their quanta known as magnons.
Magnons are known to form composite quasi-particles when they interact with other degrees of freedom.
Of particular interest is the coupled mode of magnons and phonons\cite{Weiler2011elastic,Thevenard2014prb,Gowtham2015jap} facilitated by the magnetoelastic coupling.
The magnon-phonon coupling can add functionalities to the magnon/phonon system that are otherwise absent in the uncoupled state.
For example, the coupling can increase the coherence length\cite{Delsing2019,Dumur2021} of magnons thanks to the phonons' long coherence length while the phonons can be controlled by magnetic field and acquire nonreciprocity owing to the magnons' time reversal symmetry breaking characteristics\cite{Damon1961,Jamali2013,Ishibashi2020}.

Surface acoustic waves (SAWs), a form of coherent phonons, are widely used to study the magnon-phonon coupling owing to their low damping loss and high energy density.
SAW devices\cite{White1965saw}, consisting of a pair of interdigital transducers (IDTs) and a ferromagnetic (FM) thin film element formed between the two IDTs and on a piezoelectric substrate, are commonly employed for such study.
Effects that originate from the magnon-phonon coupling in such systems have been reported recently that include, for example, the acoustic spin pumping\cite{Weiler2012fmr,Xu2018prb}, nonreciprocal propagation of SAWs\cite{Sasaki2017,kuss2020prl,Xu2020,Hernandez2020,Piyush2020,Taneto2020,Matsumoto2022,kuss2021prap}, and coherent coupling of magnon-phonon in an acoustic cavity\cite{Hatanaka2022}.

As the magnon-phonon coupling is primarily determined by the magnetoelastic coupling constant of the magnetic material, it is of great importance to quantitatively characterize its strength.
In general, evaluation of the magnetoelastic coupling constant requires mechanical tools, such as bending and/or tension applying systems\cite{Smith1963jap, Klokholm1976ieee, Lee1990prb, Betz1996apl, Gowtham2016prb, Schwienbacher2019jap}, which require special apparatus and careful calibration of the strain.
Here we propose a simple but accurate approach to determining the magnetoelastic coupling constant of a ferromagnetic layer in a thin film heterostructure formed on SAW devices.
A constant voltage across a pair of IDTs is applied to induce strain in the film.
Due to the magnetoelastic coupling, the strain causes a slight tilting in the magnetization direction, which we quantify using the planar Hall resistance of the ferromagnetic layer.
We apply this on-chip all-electrical approach to determine the magnetoelastic coupling constant of a ferromagnetic layer embedded in different heterostructures.
\begin{figure}[t]
	\centering
	\includegraphics[scale=0.125]{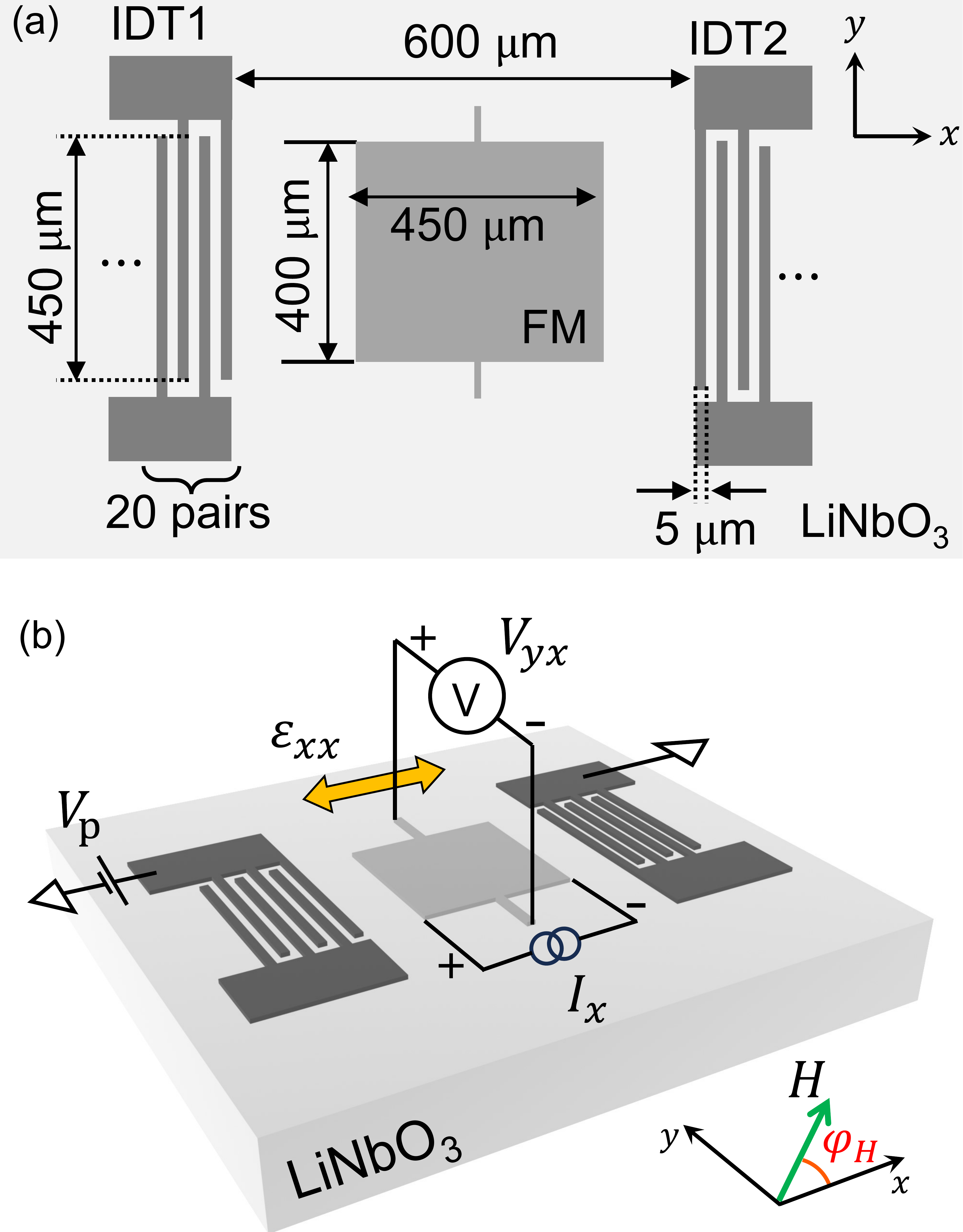}
	\caption{(a) Illustration of the device and its dimensions. (b) Schematic diagram of the experimental setup and definition of the coordinate system. The orange arrow ($\epsilon_{xx}$) represents the longitudinal strain induced by the voltage ($V_\mathrm{p}$) applied to the IDT.
		\label{fig:schematics}
	}
\end{figure}

Figure~\ref{fig:schematics} shows a schematic representation of the experimental setup, along with the coordinate system used in this study.
The devices employed here consist of a thin film heterostructure element, a pair of IDTs, and electrodes for electrical contact.
The films are deposited on piezoelectric Y$+128^\circ$-cut LiNbO$_3$ substrates using radio frequency (rf) magnetron sputtering.
The film structure is sub./X/MgO (2)/Ta (1), where X is one of the following: W (2.4)/CoFeB (1), Pt (2.4)/CoFeB (1), and MgO (2)/CoFeB (1)/W (2.4)  (thicknesses in unit of nanometers), denoted as W/CoFeB, Pt/CoFeB, and CoFeB/W, respectively, hereafter.
MgO (2)/Ta (1) serves as a capping layer to prevent oxidation of the films.
Note that the crystallinity of the W layer is different for W/CoFeB and CoFeB/W: the former is predominantly composed of the $\beta$-phase whereas the latter is a mixture of $\alpha$- and $\beta$-phases\cite{Kawada2021}.
The degree of mixture can be roughly inferred from the film resistance\cite{pai2012apl,liu2015apl}. 
For all films, the magnetic easy axis of the CoFeB layer points along the film plane.
Optical lithography and Ar ion etching are used to pattern the films into rectangular elements.
The size of the rectangle is approximately 400 $\upmu$m wide and 450 $\upmu$m long. 
Electrodes and the IDTs, composed of Ta (5)/Cu (100)/Pt (5), are patterned using optical lithography and a lift-off process.
Two IDTs are placed on the sides of the rectangular element: we denote them as IDT1 and IDT2 (see Fig.~\ref{fig:schematics}(a)).
Each IDT consists of 20 pairs of fingers, with a width and gap of approximately 5 $\upmu$m, and the length is 450 $\upmu$m.

To quantitatively assess the magnetoelastic coupling constants of the FM (here CoFeB), we study the change in the magnetization direction induced by the strain applied.
We make use of the piezoelectricity of the susbtrate to apply strain to the film.
A direct current (dc) voltage $V_\mathrm{p}$ is applied to IDT1 while IDT2 is grounded, which in turn induces a strain between the two IDTs.
The strain induced in the film is proportional to $V_\mathrm{p}$ (see supplementary material for the details).
The direction of the FM magnetization is determined by the planar Hall resistance that originates from the anisotropic magnetoresistance\cite{McGuire1975amr} and the spin Hall magnetoresistance\cite{Nakayama2013smr}.
A constant electric current ($I_{x}$) of 10 mA is supplied to the films along the $x$-axis.
The transverse voltage $V_{yx}$ across the films is measured using a voltmeter. 
The transverse resistance $R_{yx}$ is obtained from $V_{yx}$ and $I_{x}$: $R_{yx} = V_{yx}/I_{x}$.
An in-plane magnetic field of magnitude $H$ is applied during the transport measurements. 
The angle between the magnetic field and the $x$-axis is defined as $\varphi_H$.
To ensure stable measurement conditions, all measurements are performed at least one hour after the electric current begins to flow through the films.

First, we describe the change in $R_{yx}$ induced by the strain. We denote the angle between the equilibrium magnetization direction without the strain and the $x$-axis as $\varphi$.
(The magnetization lies within the film ($xy$) plane.)
The strain-induced change in the magnetization direction $\delta \varphi$ can be expressed as follows\cite{Weiler2012Ni,Kawada2021}:
\begin{equation}
	\centering
	\begin{aligned}
		\delta \varphi = \delta \varphi_0 \sin 2\varphi_H,
		\label{eq:deltaphi}
	\end{aligned}
\end{equation}
where the amplitude $\delta \varphi_0$ is represented by
\begin{equation}
	\centering
	\begin{aligned}
		\delta \varphi_0 = \frac{H_\mathrm{ME}}{H}.
		\label{eq:deltaphi0}
	\end{aligned}
\end{equation}
$H_\mathrm{ME}$ represents the effective magnetic field induced by strain and is given as
\begin{equation}
	\centering
	\begin{aligned}
		H_\mathrm{ME} = -\frac{b\epsilon_{xx}(V_\mathrm{p})}{M_\mathrm{S}}.
		\label{eq:hme}
	\end{aligned}
\end{equation}
$\epsilon_{xx}(V_\mathrm{p})$ is the strain along the $x$-axis, $\ M_\mathrm{S}$ and $b$ are the saturation magnetization and the magnetoelastic coupling constant of the FM layer, respectively. 
Upon application of $V_\mathrm{p}$ to IDT1, $R_{yx}$ under strain is expressed as
\begin{equation}
	\centering
	\begin{aligned}
		R_{yx}(V_\mathrm{p})&=R_{yx}^{2\varphi} \cdot \frac{1}{2}\sin 2\qty(\varphi_H+\delta \varphi) \\
		&\approx R_{yx}^{2\varphi} \cdot \frac{1}{2}\sin 2\varphi_H+R_{yx}^{4\varphi} \cdot \frac{1}{2}\sin 4\varphi_H,
		\label{eq:smr_strain_xy}
	\end{aligned}
\end{equation}
where in the second line, we defined 
\begin{equation}
	\centering
	\begin{aligned}
		R_{yx}^{4\varphi} \equiv R_{yx}^{2\varphi} \delta \varphi_0 = - R_{yx}^{2\varphi} \frac{b}{H M_\mathrm{S}} \epsilon_{xx}(V_\mathrm{p}),
		\label{eq:rxy4phi}
	\end{aligned}
\end{equation}
using Eqs.~(\ref{eq:deltaphi0}) and (\ref{eq:hme}). It is evident that $R_{yx}^{4\varphi}$ contains information on the magnetoelastic coupling while $R_{yx}^{2\varphi}$ represents the magnitude of the planar Hall resistance at $V_\mathrm{p}=0$ V. 
Note that Eq.~(\ref{eq:smr_strain_xy}) holds under the condition $\delta \varphi_0 \ll 1$, which is satisfied in the present study.

\begin{figure}[t]
	\centering
	\begin{minipage}{1.0\hsize}
		\centering
		\includegraphics[scale=0.15]{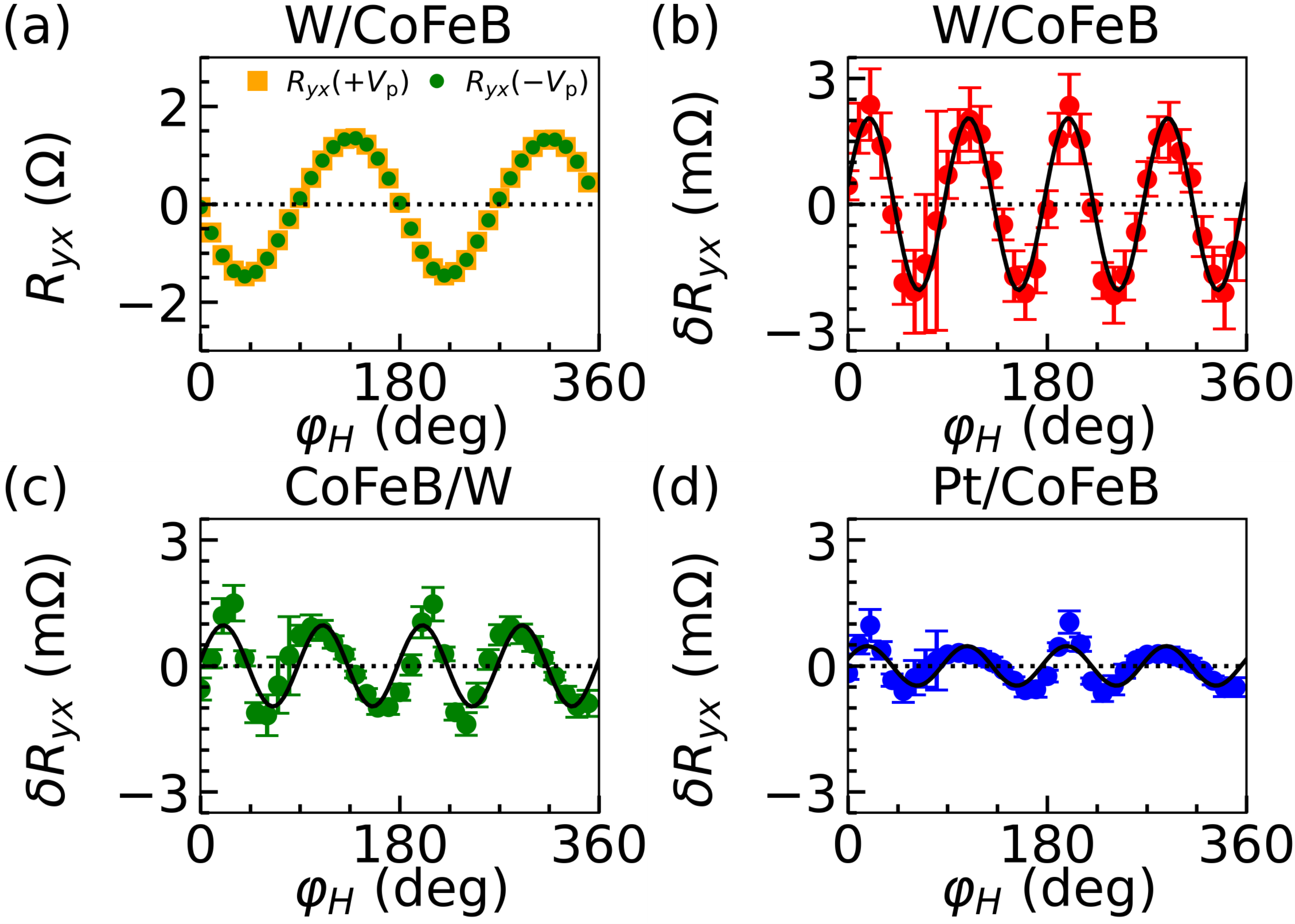}
	\end{minipage}

	\caption{
		(a) Magnetic field angle ($\varphi_H$) dependence of ${R}_{yx}$ for W/CoFeB (a). The orange squares (purple dots) represent the results when a dc voltage $V_\mathrm{p}$ of $+210$ V ($-210$ V) is applied to IDT1.  (b-d) $\delta R_{yx}$ plotted against $\varphi_H$ for W/CoFeB (b), CoFeB/W (c), and Pt/CoFeB (d). The magnetic field magnitude $H$ is fixed to $\sim$7 mT and $|V_\mathrm{p}| = 210$ V. The error bars are standard deviation of the repeated measurements. The black lines in (b-d) show fit to the data with Eq.~(\ref{eq:4theta_xy}).
		\label{fig:4theta}
	}
\end{figure}

The measurement results of $R_{yx}(V_\mathrm{p})$ for W/CoFeB with $V_\mathrm{p}=\pm210$ V are plotted against $\varphi_H$ in Fig.~\ref{fig:4theta}(a).
Since the change in $R_{yx}$ caused by the strain [$R_{yx}^{4\varphi}$ in Eq.~(\ref{eq:smr_strain_xy})] is significantly smaller than $R_{yx}^{2\varphi}$, it is difficult to resolve its effect from such plot.
We therefore process data obtained from $R_{yx}(+V_\mathrm{p})$ and $R_{yx}(-V_\mathrm{p})$.
Noting that $\epsilon_{xx}(-V_\mathrm{p})=-\epsilon_{xx}(+V_\mathrm{p})$, we define the following quantities:
\begin{equation}
	\centering
	\begin{aligned}
		\overline{R}_{yx} \equiv \frac{R_{yx}(+V_\mathrm{p})+R_{yx}(-V_\mathrm{p})}{2}
		\sim R_{yx}^{2\varphi} \cdot \frac{1}{2}\sin 2\varphi_H,\\
		\delta R_{yx}(V_\mathrm{p}) \equiv \frac{R_{yx}(+V_\mathrm{p})-R_{yx}(-V_\mathrm{p})}{2}
		\sim R_{yx}^{4\varphi} \cdot \frac{1}{2}\sin 4\varphi_H.
		\label{eq:4theta_xy}
	\end{aligned}
\end{equation}
Thus $R_{yx}^{2\varphi}$ and $R_{yx}^{4\varphi}$ can be obtained from $\overline{R}_{yx}$ and $\delta R_{yx}(V_\mathrm{p})$, respectively.

\begin{figure}[t]
	\centering
	\begin{minipage}{1.0\hsize}
		\centering
		\includegraphics[scale=0.15]{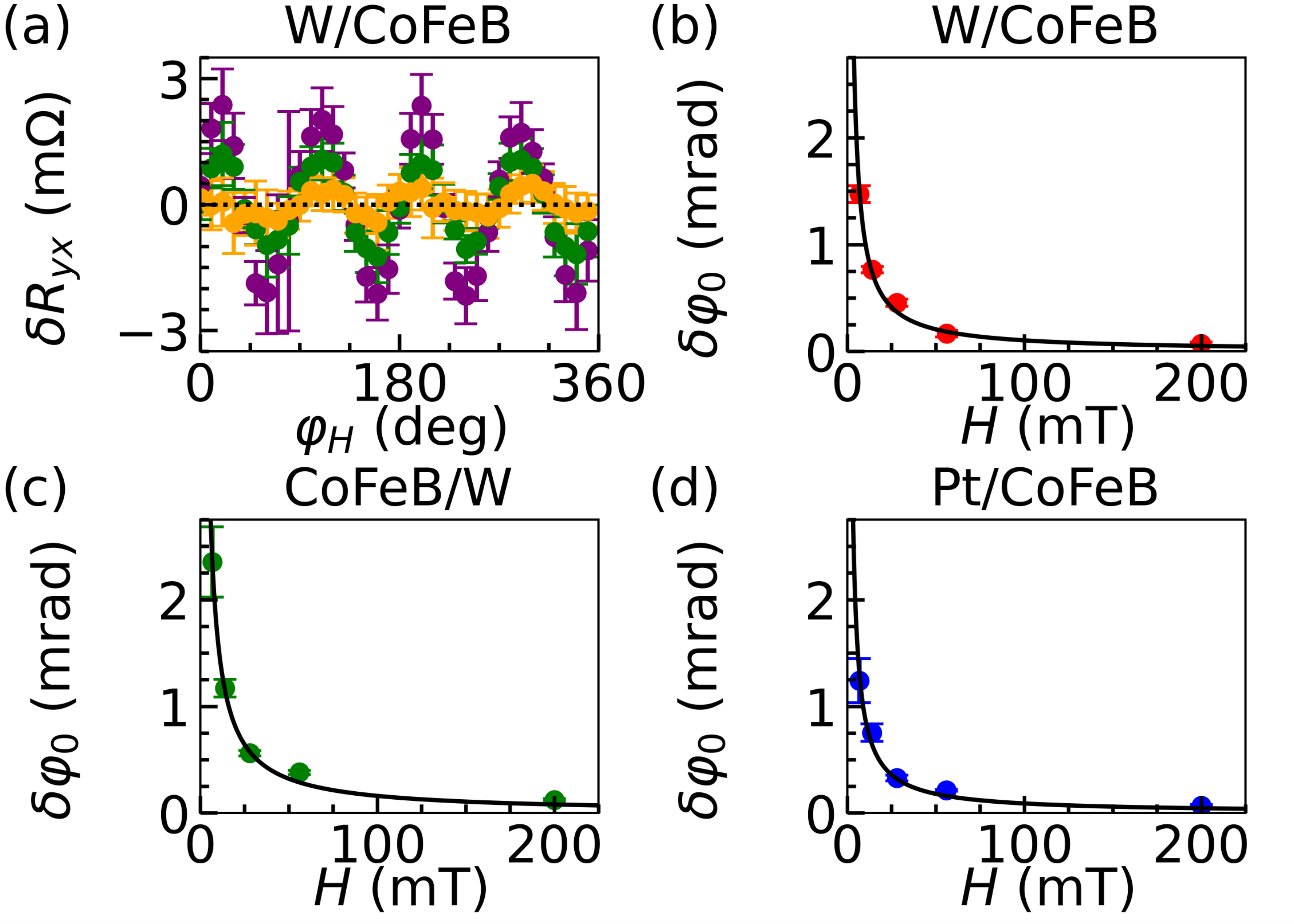}
	\end{minipage}
	\caption{(a) The field angle ($\varphi_H$) dependence of $\delta R_{yx}$ for W/CoFeB measured under various field magnitudes ($H$). Purple, green and orange points are for $H\sim$ 7 mT, 14 mT, and 56 mT, respectively.$|V_\mathrm{p}| = 210$ V is applied. The offset resistance is subtracted from the plotted data. The error bars indicate standard deviation of the repeated measurements. (b-d) The field magnitude ($H$) dependence of $\delta \varphi_0$ for W/CoFeB (b), CoFeB/W (c), and Pt/CoFeB (d) with $|V_\mathrm{p}|=210$ V. The error bars represent fitting errors. The black lines show fit to the data with Eq.~(\ref{eq:deltaphi0}).
		\label{fig:field}
	}
\end{figure}

Figures~\ref{fig:4theta}(b-d) show the $\varphi_H$ dependence of $\delta R_{yx}$ for W/CoFeB, CoFeB/W, and Pt/CoFeB with $H\sim7$ mT and $|V_\mathrm{p}|=210$ V.
As evident, $\delta R_{yx}$ sinusoidally changes with $\varphi_H$ in a period of 90$^\circ$ for all the films.
We fit the data using Eq.~(\ref{eq:4theta_xy}) to extract $R_{yx}^{4\varphi}$.
Deviation from the $\sin 4\varphi_H$ dependence are found for Pt/CoFeB and to a lesser extent, in CoFeB/W.
We believe this is due to the magnetic anisotropy of the FM layer.
Note that the signal strength ($R_{yx}^{4\varphi}$) depends on the planar Hall resistance $R_{yx}^{2\varphi}$ of the samples: see Eq.~(\ref{eq:rxy4phi}).
The planar Hall resistance ratio, $R_{yx}^{2\varphi}$ divided by the longitudinal resistance, is $0.5\% \sim 1$\% for the samples studied here. For a single layer CoFeB films, the planar Hall resistance ratio is less than 0.1\%, which may require longer measurement time to obtain processable data.


According to Eq.~(\ref{eq:rxy4phi}), $R_{yx}^{4\varphi}$ must scale with the inverse of $H$.
We therefore study the magnetic field magnitude ($H$) dependence of $\delta R_{yx}$ (and thereby the $H$ dependence of $R_{yx}^{4\varphi}$).
Figure~\ref{fig:field}(a) shows the $\varphi_H$ dependence of $\delta R_{yx}$ for W/CoFeB measured with various $H$.
Here $\abs{V_\mathrm{p}}$ is fixed to 210 V.
The results show that the amplitude of the $\sin 4\varphi_H$ variation indeed decreases with increasing $H$.
We thus fit the $\varphi_H$ dependence of $\overline{R}_{yx}$ (not shown here) and $\delta R_{yx}$ with Eq.~(\ref{eq:4theta_xy}) to obtain $R_{yx}^{2\varphi}$ and $R_{yx}^{4\varphi}$, respectively.
Subsequently, we normalize $R_{yx}^{4\varphi}$ by dividing it with $R_{yx}^{2\varphi}$ to estimate $\delta \varphi_0$: see Eq.~(\ref{eq:rxy4phi}).
The $H$ dependence of $\delta \varphi_0$ for W/CoFeB, CoFeB/W, and Pt/CoFeB are plotted in Figs.~\ref{fig:field}(b-d).
We fit the data in Figs.~\ref{fig:field}(b-d) with Eq.~(\ref{eq:deltaphi0}) to determine the magnetoelastic effective field $H_\mathrm{ME}$.
The fitting results are shown by the solid lines in Figs.~\ref{fig:field}(b-d), which show good agreement with the experimental results.

\begin{table}[t]
	\caption{Summary of the parameters obtained from the experiments for W/CoFeB, CoFeB/W, and Pt/CoFeB. Magnetoelastic effective field: $H_\mathrm{ME}$, saturation magnetization: $M_\mathrm{S}$, thickness of the magnetic dead layer: $t_\mathrm{d}$, magnetoelastic energy with a piezoelectric strain of $|V_\mathrm{p}| = 210$ V: $b\epsilon_{xx}$. Errors represent the fitting errors.}
	\label{tab:fit_results}
	\begin{tabular}{ccccc}
	\hline \hline
	Film & $H_\mathrm{ME}$ ($\upmu$T) & $M_\mathrm{S}$ (MA/m) & $t_\mathrm{d}$ (nm) & $b\epsilon_{xx}$ ($\upmu$J/cm$^3$)\\
	\hline
	W/CoFeB & $10.4 \pm 0.3$ & $1.31 \pm 0.06$ & $0.3 \pm 0.1$ & $14 \pm 1$\\
	CoFeB/W & $16.1 \pm 0.3$ & $1.15 \pm 0.05$ & $0.4 \pm 0.1$ & $19 \pm 1$\\
	Pt/CoFeB & $9.0 \pm 0.4$ & $1.12 \pm 0.06$ & $-0.5 \pm 0.2$ & $10 \pm 1$\\
	\hline\hline
	  \end{tabular}
\end{table}

The parameters obtained from the fitting are listed in Table~\ref{tab:fit_results}.
We find $H_\mathrm{ME}$ of the order of 10 $\upmu$T for all samples.
To extract the magnetoelastic coupling constant $b$ from $H_\mathrm{ME}$, the saturation magnetization $M_\mathrm{S}$ of CoFeB must be known for each film structure.
Magnetic moment of the samples are measured using a vibrating sample magnetometer.
The thickness of the CoFeB layer is varied to extract $M_\mathrm{S}$.
Specifically, films composed of sub./X/MgO (2)/Ta (1) with X=W (2.4)/CoFeB ($d$), Pt (2.4)/CoFeB ($d$), and MgO (2)/CoFeB ($d$)/W (2.4) (thickness in unit of nanometers) are grown on Si substrates coated with 100-nm-thick silicon oxide (SiO$_x$) under the same sputtering condition as W/CoFeB, Pt/CoFeB, and CoFeB/W.
The CoFeB thickness $d$ ranges from 1 nm to 2.5 nm.

\begin{figure}[b]
	\centering
	\begin{minipage}{1.0\hsize}
		\centering
		\includegraphics[scale=0.135]{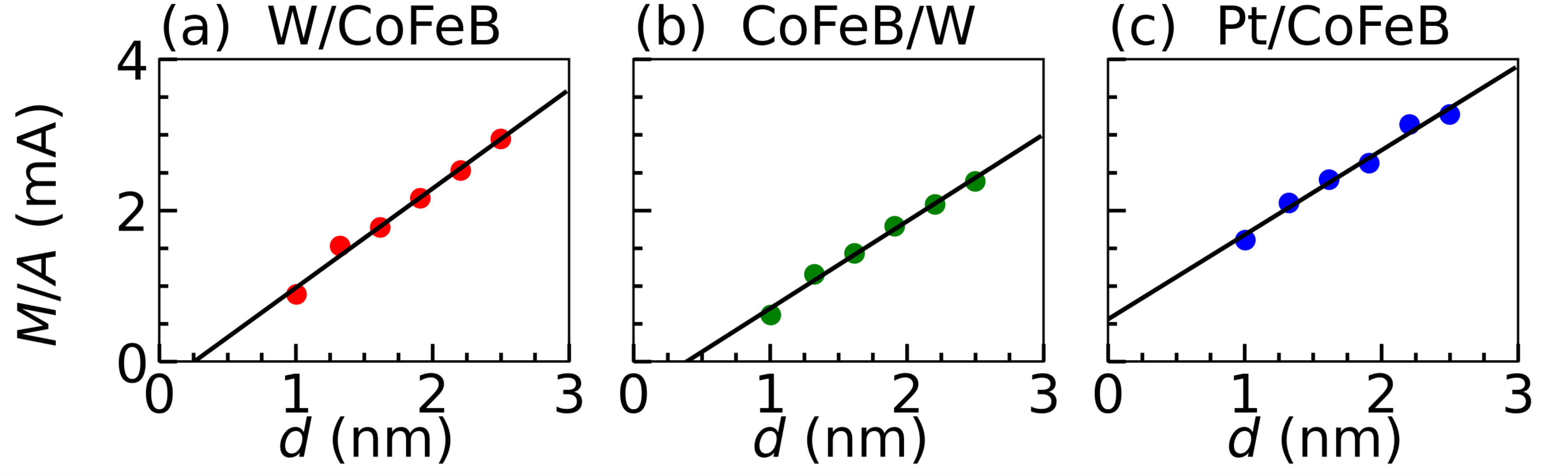}
	\end{minipage}
	\caption{(a-c) CoFeB thickness ($d$) dependence of the magnetic moment $M$ per film area $A$ for W/CoFeB (a), CoFeB/W (b), and Pt/CoFeB (c). The black straight lines show linear fit to the data. 
		\label{fig:ms_mat}
	}
\end{figure}

Figures~\ref{fig:ms_mat}(a-c) show the $d$ dependence of the magnetic moment ($M$) divided by the film area ($A$) for W/CoFeB, CoFeB/W, and Pt/CoFeB (on the SiO$_x$ substrates). $M/A$ linearly scales with $d$ for all structures.
The data is fitted with a linear function: the intercept to the $x$-axis and the slope give the magnetic dead layer thickness $t_\mathrm{d}$ and the saturation magnetization $M_\mathrm{S}$, respectively. 
The obtained values are summarized for each structure in Table~\ref{tab:fit_results}.
$t_\mathrm{d}$ is consistent with previous studies\cite{Jang2010jap,Sinha2013apl}. The negative $t_\mathrm{d}$ of Pt/CoFeB may originate from the proximity-induced moment of Pt \cite{Huang2012prl,Fan2014nc,ueno2015scirep}. 
The obtained $M_\mathrm{S}$ is close to that reported in Ref.~\cite{Gowtham2016prb}. $M_\mathrm{S}$ for W/CoFeB is slightly larger than the others, which may originate from a stronger boron absorbing effect of W during deposition of CoFeB. 

Using Eq.~(\ref{eq:hme}), we extract the product of the magnetoelastic coupling constant $b$ and the applied strain $\epsilon_{xx}$ using the parameters shown in Table~\ref{tab:fit_results}.
As evident, $b\epsilon_{xx}$ varies among the three structures studied. 
Assuming that the strain $\epsilon_{xx}$ applied to the samples is the same, these results suggest that $b$ is dependent on the layer adjacent to the FM layer.
We infer that this is due to the difference in the elastic properties of the bilayers. 
Using the magnetostriction constant $\lambda_s$ and the shear modulus $G$, $b$ is expressed as $-3\lambda_s G$\cite{Betz1996apl}. 
While $\lambda_s$ is the same for a given material (CoFeB), $G$ is set by all layers that form the heterostructure.
Previous studies reported that $G$ for bulk Pt (fcc) is 2.4 times smaller than that for bulk W (bcc)\cite{Darling1966}. 
Studies have shown that $G$ for amorphous metals is about two-thirds of that for crystalline counterparts\cite{Knuyt1986jpf}.
Such reported values of $G$ for the non-magnetic metals suggest that $G$ is the largest for CoFeB/W and the smallest for Pt/CoFeB.
This trend is consistent with the results shown in  Table~\ref{tab:fit_results}.

In summary, we have developed an approach that allows determination of the magnetoelastic coupling constant of magnetic thin film heterostrucutres patterned on a SAW device.
Owing to the piezoelectricity of the substrate, a dc voltage applied to a pair of IDT-shaped electrodes induces strain in the film formed between the IDTs.
The strain causes the direction of the magnetization to tilt due to the magnetoelastic coupling.
The change in the magnetization direction is probed using the planar Hall resistance of the film.
From the transport measurements, we determine the magnetoelastic coupling constant of the CoFeB layer in various thin film heterostructures. 
We find the magnetoelastic coupling constant is dependent on the film stacking, which is primarily attributed to the difference in the elastic properties (shear modulus) of the other layers in the heterostructure.
These results thus show the importance of the film stacking on the magnetoelastic coupling constant of a magnetic layer in thin film heterostructures.
The method presented in this paper provides a versatile approach to determine the magneto-elastic coupling constant, a key parameter for spin mechantronics\cite{Matsuo2017spinmec} and magnomechanics\cite{Delsing2019,Barman2021}.

See supplementary material for the $V_\mathrm{p}$ dependence of $\delta R_{yx}$ and $\delta \varphi_0$.

\section{Acknowledgements}
This work was partly supported by JSPS KAKENHI (Grant Numbers 20J21915, 23KJ1419, and 23H05463), MEXT Initiative to Establish Next-generation Novel Integrated Circuits Centers (X-NICS).

\bibliography{reference_081123}

\end{document}